# Assessment of Practical Smart Gateway Diversity Based on Multi-Site Measurements in Q/V band

Spiros Ventouras, Pantelis-Daniel Arapoglou

*Abstract*— Next generations of High Throughput Satellite (HTS) systems operating a Q/V band network of gateways (GWs) for the feeder link, need to apply some form of Smart Gateway Diversity (SGD) to ensure they meet the demanding availability targets. However, most of the published research on SGD usually resorts to ideal assumptions that are not realistic for an operational system. To evaluate in more depth the performance of practical SGD, this paper exploits for the first time multi-site propagation measurements for a network of six measurement sites (GWs) at 40 GHz. Based on this, it studies the performance of SGD in terms of feeder availability and number of switches between GWs taking into account the climatic region, the switching processing delay, the clustering of GWs, and the impact of downlink versus uplink frequency in Q/V band. Along the course, the interplay between propagation aspects and SGD operational aspects is highlighted.

## I. Introduction

For exploiting Extremely High Frequency (EHF) bands beyond 10 GHz in the RF feeder links of High Throughput Satellite (HTS) networks, it is necessary to rely on some sort of Smart Gateway Diversity (SGD) [1], [2], [3] since this technique ensures the high availability targets required. The spectrum needs resulting from HTS systems lead to already deploying multiple gateway stations (GWs) for nominal service. This is the reason why, in recent years, the cost of the HTS ground segment is absorbing a much larger part of the overall mission cost, a source of growing concern. SGD enables the cooperation between the already existing GWs and ensures the target availability by potentially adding only a few extra resources (GWs) for cold redundancy. This is in sharp contrast with traditional site diversity where each ground station needs to be paired with a redundant one, which leads to a doubling of the number of GWs.

SGD exploits the geographical spatial diversity provided by interconnecting multiple GW stations in different feeder beams in order to increase the ground network availability. Two main flavours of SGD have been proposed [1]: the N active GWs and the N active + P redundant GWs. They both operate under the same fundamental premise of re-routing data from a faded GW to a different GW site of another feeder beam[1]. Their difference is that, in the first scheme (N), each GW is oversized in terms of resources to handle part of the traffic of a faded GW in the network, whereas, in the second scheme (N+P), the N nominal GWs are fully loaded so that when one falls in outage, one of the P redundant ones completely takes over its traffic. Both techniques require substantial ground-based connectivity and the associated spacecraft RF routing mechanisms; as such the choice of SGD has an impact on the satellite payload architecture [4], [5]. For ensuring very high feeder link availabilities (99.9% or even 99.99%), an active GW under heavy fading needs to switch to another one. Such a switching event[2] risks the loss of end user demodulator synchronization due to the land-based delay in re-routing data between GWs.

It is clear that the spatial and temporal characteristics of all the feeder propagation channels are driving the performance of the SGD technique and that its effectiveness in the system performance depends on both. Hitherto, with the exception of few recent works (e.g. [6], [7]) most of the contributions on the topic of SGD have made ideal assumptions, such as perfect de-correlation of fading amid the GW locations, ideal knowledge of the fading and ideal prediction of its evolution, as well as instantaneous switching between the GW sites. In practice, the practicalities in operating SGD may play a detrimental role in the performance of SGD in terms of offered availability. Such practicalities include the switching/handover dynamics under practical constraints (e.g.

This paragraph of the first footnote will contain the date on which you submitted your paper for review.. This work was partially funded by ESA Contract "Post Processing of Propagation Measurements for Smart GW Applications" and EPSRC Grant EP/P0246.29/1 "Optimising resources in future heterogeneous millimeter wave communication systems" The emulations were derived using a prototype SW tool developed within the project "Developing Software Interface for SatCom Industrial Users of RAL Space Facilities " funded by STFC under the scheme Commercial Pump Prime Fund.

S. Ventouras is with UKRI-STFC Rutherford Appleton Laboratory, Harwell Campus, Didcot,OX11 0QX, UK (e-mail: spiros.ventouras@stfc.ac.uk).

P-D.Arapoglou is with ESA/ESTEC (e-mail:pantelis-daniel.arapoglou@esa.int).
.

---

[1] Instead, with traditional site diversity, a secondary antenna is placed within the same beam kilometers away from the faded GW site.

[2] Switching and handover are interchangeably used in the rest of the paper.

the frequency of GW switching/handover or the criterion when to switch), the switching characteristics between GWs and the related channel estimation.

There is a clear gap of such studies and models due also to the fact that existing concurrent and synchronized time series of attenuation measured from a network of experimental sites (such as the ALPHASAT propagation experiment [8]) have never been analyzed towards this goal. The existing propagation models (ITU-R Study Group 3 Recommendations) are statistical prediction models aiming mainly at the specification of the fade margin or at the description of the fade dynamics (e.g. fade duration, fade slope etc.) both inconclusive in determining either the performance of the real systems or how to optimize its operation.

In this paper, the authors try to fill this gap by emulating for the first time the performance of the SGD technique employing real multi-site data (measurements) of the propagation channels in Q/V band. For easier reading of the results, the N+P variant of SGD is adopted throughout the paper, although this is expected to have small impact on the main results and conclusions. The main novelty of the paper is that the emulations are based on measurements of the Q-band (40 GHz) propagation signal transmitted by the Aldo Paraboni payload (TDP5) from the Alphasat satellite. Although a specific set of measurements should not be directly used for designing any HTS system, the exercise offers great insights into the SGD technique.

The measurements are the results of a joint propagation campaign at pan-European level, which was coordinated across five countries [9]. Although measured data have been used in the past to assess aspects of SGD [6], in this paper six different sites is considered, a number which in terms of GWs of a feeder network would already match a medium sized HTS system of about 100 Gbps. Section II presents the experimental sites of the measurements and the climatic patterns which were found to be closely related to the SGD technique. Section III describes SGD key parameters and Section IV reports on the results of the emulations for different system requirements and system scenarios including among others the impact of switching delay, of payload connectivity restrictions as well as the impact of carrier frequency (downlink frequency versus uplink frequency). The conclusions are drawn in Section V.

## II. EXPERIMENTAL AND PROPAGATION CHARACTERISTICS

The data used in the SGD emulations are collected using the Q-band experimental configuration of ASALACA experiment [9] at the following locations: {UK: (Chilbolton, Chilton), Spain: (Vigo), Portugal: (Aveiro) and Greece: (Athens, Lavrion)} with corresponding elevation angles {UK: (26.40°,26.07°), Spain: (30.60°), Portugal:(31.80°), Greece: (45.97°,46.26°)}. The observation period of the used measurements is of 12 months: from 1st January 2018 to 31th of December 2018 and the concurrent valid data availability for all locations is greater than 90% which guarantees the reliability of the results. The distances between Chilbolton-

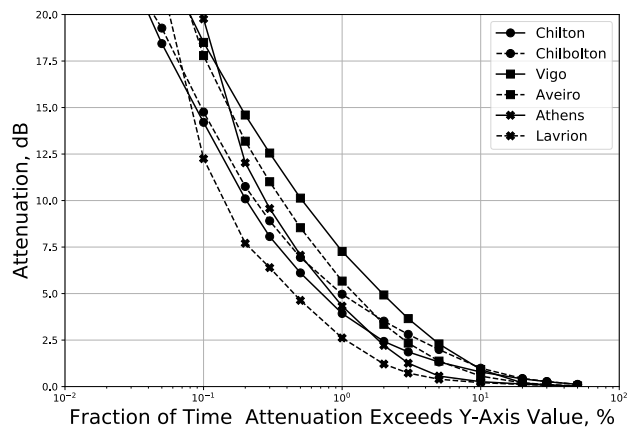

Fig. 1. Annual attenuation statistics measured at the 6 locations at 40 GHz.

TABLE I
MEASURED FADING TIME OF 40GHZ EXCESS ATTENUATION AT CHILTON, CHILBOLTON; VIGO, AVEIRO; ATHENS, LAVRION.

| Location | dB | Fading Time (%) | Fade Time (minutes) | No. of Fades | Mean Duration (sec) |
|---|---|---|---|---|---|
| Chilton | 5 | 0.695 | 3652.6 | 5726 | 36.73 |
|  | 10 | 0.203 | 1069.0 | 1802 | 34.16 |
| Chilbolton | 5 | 0.985 | 5178.1 | 8648 | 34.77 |
|  | 10 | 0.234 | 1229.0 | 1689 | 42.26 |
| Vigo | 5 | 1.955 | 10276.0 | 12264 | 48.60 |
|  | 10 | 0.515 | 2707.3 | 3393 | 46.28 |
| Aveiro | 5 | 1.199 | 6301.8 | 7249 | 50.57 |
|  | 10 | 0.367 | 1930.5 | 2122 | 52.92 |
| Athens | 5 | 0.810 | 4259.4 | 4241 | 58.14 |
|  | 10 | 0.278 | 1459.6 | 1249 | 67.66 |
| Lavrion | 5 | 0.446 | 2346.0 | 2155 | 63.29 |
|  | 10 | 0.129 | 677.0 | 453 | 86.89 |

Chilton, Vigo-Aveiro and Athens-Lavrion are 47.9Km, 173.3Km, 36.31Km respectively. For the other pairs of the experimental sites the distances are longer than 1000Km.

The terminals of the measurement network are time synchronized (using GPS) to ensure the applicability of measurements for temporal and spatial correlation studies and measure excess attenuation, i.e. the variation of signal that can be ascribed mainly to rain and cloud attenuation. Furthermore, the initially derived attenuation time series were further processes to make sure they are harmonized to a common time resolution of 1 second.

This achievable coverage area is representative of the fringe of a typical satellite covering Europe and encompassing several climatic zones (from north Atlantic to the southern Mediterranean areas). The experimental sites can be clustered into 3 distinct climatic regions in terms of radio propagation: Southern UK (Chilton, Chilbolton); Spain–Portugal (Vigo, Aveiro); Greece (Athens, Lavrion). Hence, the experimental sites encompass both macro- and mid-scale diversity effects. This allows us to extrapolate also some conclusions for a feeder network within even a higher number of GWs across Europe, with a denser distribution. Fig. 1 shows the annual attenuation statistics measured at the 6 locations at 39.4 GHz[3], whereas Table I lists the measured fading time and total number of fades for fades exceeding 5 dB and 10 dB. The choice of these fading margins will be motivated in the following sections. Fig. 2 shows the distribution of fading time versus the fade duration at 10 dB. The distinct propagation characteristics of the three regions are clearly visible. In general, it seems that the sites in Spain, Portugal and Greece experience severer fades than in Southern UK. Regarding the number of fades, the largest numbers observed in Vigo and the lowest in Greece. This is reflected in the mean fade duration values listed in Table I (i.e. fading time over total number of fades at a given threshold); the longest mean fade duration values observed in Greece and the shortest in UK.

---
[3] For brevity, we will be rounding the beacon frequency carrier to 40 GHz.

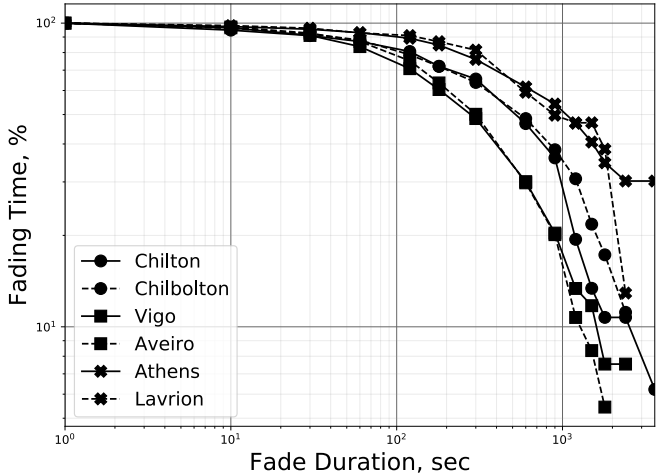

Fig. 2. Measured Fading time due to fades exceeding 10dB with duration longer than x-axis value for all the locations. The fading time is normalised to the outage at 10dB.

Regarding the spatial characteristics of the radio channels, Table II lists the joint statistics, that is the time the attenuation at both sites (GWs in the following assuming SGD) exceeds 5 dB and 10 dB. Of interest to the SGD technique is the fact that the time an attenuation threshold is exceeded on both GW links of a pair is not strictly related to the distance between the two GWs, but rather to meteorological characteristics of the GW locations (e.g. weather front). That is, shorter distances do not always imply higher exceedance time. As expected, for the pairs with GWs in the same climatic region the joint exceedance time is much higher than the other GW pairs, i.e. there is a stronger correlation of the propagation impairments. This is clearly illustrated in Fig. 3 which depicts annual joint attenuation statistics for 6 pairs of GWs – 3 pairs with GWs from the same climatic region. For example, the joint attenuation statistics between Vigo-Aveiro is comparable to joint statistics between Chilton-Chilbolton and Athens-Lavrion regardless the long separation of 173.3Km between Vigo and Aveiro.

III. SMART GATEWAY DIVERSITY METRICS AND PARAMETERS

The primary metrics for the SGD assessment is the achievable feeder network availability and number of required switches over a period of interest T (e.g. a year, a day). At any instant, the feeder network is available if N (or more) GWs out of the N+P GWs are not in outage. Over a period of interest T, the feeder network availability is defined as the percentage of T the system is available.

TABLE II
MEASURED JOINT EXCEEDANCE TIME FOR THE 15 GW LINK PAIRS AT 5 DB AND 10 DB.

| GW pair | Outage in minutes | |
|---|---|---|
| | 5dB | 10dB |
| *Chilton-Chilbolton* | 1303.0 | 149.2 |
| Chilton-Vigo | 204.6 | 20.6 |
| Chilton-Aveiro | 61.5 | 8.2 |
| Chilton-Athens | 7.6 | 0 |
| Chilton-Lavrion | 4.1 | 0 |
| Chilbolton-Vigo | 350.5 | 41.7 |
| Chilbolton-Aveiro | 100.4 | 5.0 |
| Chilbolton-Athens | 14.3 | 0 |
| Chilbolton-Lavrion | 11.4 | 0.1 |
| *Vigo-Aveiro* | 735.9 | 66.8 |
| Vigo-Athens | 77.1 | 4.1 |
| Vigo-Lavrion | 53.3 | 4.8 |
| Aveiro-Athens | 94.7 | 10.0 |
| Aveiro-Lavrion | 72.2 | 4.5 |
| *Athens-Lavrion* | 675.5 | 96.1 |

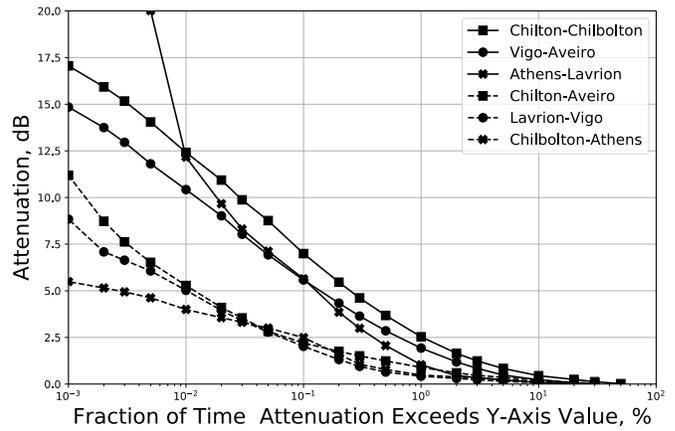

Fig. 3. Measured annual fraction of time for which attenuation at both links exceeds the Y-Axis value for GWs located in the same or different climatic regions.

A Smart Gateway N+P Network Configuration is primarily determined by the required number of GWs (active and redundant) and the geographic location, the operational frequency and the available Fade Margin (FM) of each GW. When the attenuation experienced by a GW exceeds the FM, the GW is in outage. It has to be noted here that the attenuation levels tolerated by the GW (FM) can be the result of several margins built into the GW's link budget, such as oversizing of HPA and/or reflector size. Out of this oversizing, the HPA also manages to apply uplink power control for compensating the uplink fading and keeping under control the co-channel interference between the GWs.

Because of the required switching, critical parameters of the system are: a) the Site Switching Threshold (SST) and b) the switching process delay, i.e. the time required for a switching to be complete. The SST is not necessarily equal to GW FM.

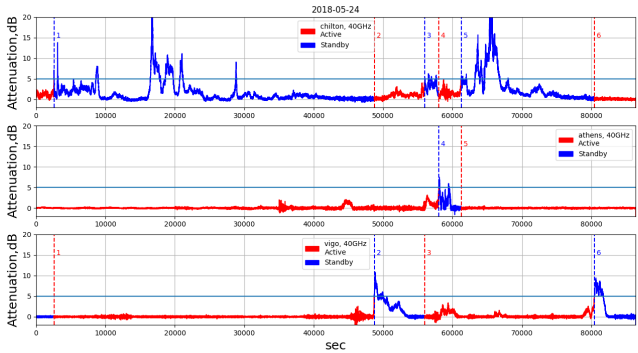

Fig. 4. Example of switching mechanism. Attenuation time series are shown for Chilton, Athens and Vigo on 24 May 2018. A GW location is active when the time series is red, it is in standby when it is blue. The number of switches at Chilton, Athens and Vigo is 6, 2 and 4 respectively whereas the network number of switches is 6=(12/2).

For example, for a tolerated excess attenuation margin FM, the network may initiate the switching process for values smaller than FM in order to avoid network outages. The sizing of this hysteresis margin is a trade-off between availability and number of switches.

The switching mechanism adopted for the emulations presented herein is as follows: under adverse propagation conditions, an active $GW_1$ switches its traffic to another redundant $GW_2$ whenever its SST is exceeded for the first time. Then, $GW_1$ becomes redundant (or standby) until the propagation conditions at another $GW_k$ require switching to $GW_1$. A switch involves two stations: a) the Active GW which becomes Standby and b) the Standby GW which becomes Active. Therefore, the total number of Network switches is the half of the sum of the switches of each GW. Fig. 4 shows an example of the switching mechanism for a (2+1) network with GWs at Chilton, Athens and Vigo with SST=5dB on 24th of May 2018.

In a N active + P redundant GWs network with P=1, when an active GW crosses the SST there is only one GW which might be used. However, for P>=2, there might be more than one GWs available and a decision has to be made which of the standby GWs will become active. The analysis of the data showed that the network availability is independent of the selection of the GW, however, the number of required switches to achieve this availability does depend on the GW choice. Therefore, for an unbiased statistical assessment of SGD performance in the emulations, a GW is selected randomly from the available GWs. Similarly, the initial selection of the standby GWs does not have an impact on the statistical assessment of SGD performance.

## IV. SGD Assessment

The 6 experimental sites enable the assessment of a 4+2 SGD technique. In addition, in order to assess the SGD technique if only one GW within the same cluster can be selected in the event of an outage, the total number of (4+2) GWs is split into two sub-sets, each containing (2+1) GWs. This situation may result from connectivity restrictions between GWs and user beams at the satellite payload or between GWs on ground. The assessment is performed for ideal switching (i.e. without considering any switching process delay), as well as for a switching process delay of 2 seconds and 30 seconds. A common dimensioning has been adopted for each GW and, hence, a common SST which is either 5 dB or 10 dB. Although this leads to a different single GW availability in each location, it is a practical hypothesis based on the assumption that all GWs will be equipped with the same equipment (antennas, HPAs). In any case, the results can be easily expanded for any choice of frequency and SST per GW.

TABLE III
MEASURED ANNUAL AVAILABILITY FOR ALL COMBINATIONS OF 4-GW NETWORKS WITHOUT SGD

| GW from Region 1 | GW from Region 2 | GW from Region 3 | GW at | Network annual availability (%) | |
|---|---|---|---|---|---|
| | | | | 5dB | 10dB |
| Chilton | Vigo | Athens | Chilbolton | 95.8664 | 98.7911 |
| Chilton | Vigo | Lavrion | Chilbolton | 96.2361 | 98.9402 |
| Chilton | Aveiro | Athens | Chilbolton | 96.5357 | 98.9258 |
| Chilton | Aveiro | Lavrion | Chilbolton | 96.9249 | 99.0769 |
| Average Availability of Networks including Chilton and Chilbolton GWs | | | | 96.3908 | 98.9335 |
| Chilton | Vigo | Lavrion | Aveiro | 95.8920 | 98.7819 |
| Chilbolton | Vigo | Athens | Aveiro | 95.3497 | 98.6269 |
| Chilbolton | Vigo | Lavrion | Aveiro | 95.7292 | 98.7815 |
| Chilton | Vigo | Athens | Aveiro | 95.5146 | 98.6300 |
| Average Availability of Networks including Vigo and Aveiro GWs | | | | 95.6214 | 98.7051 |
| Chilton | Vigo | Athens | Lavrion | 96.2305 | 98.8675 |
| Chilbolton | Vigo | Athens | Lavrion | 96.0725 | 98.8715 |
| Chilton | Aveiro | Athens | Lavrion | 96.9567 | 99.0112 |
| Chilbolton | Aveiro | Athens | Lavrion | 96.7224 | 99.0017 |
| Average Availability of Networks including Athens and Lavrion GWs | | | | 96.4956 | 98.9378 |
| Average of all 4+0 Networks | | | | 96.1692 | 98.8586 |

### A. (4+0) without SGD Deployment

In general, in a N+P SGD network the number N of active GWs is determined by the required HTS system capacity whereas the deployment of P redundant GWs is to increase the network availability. In this section we discuss the availability of feeder networks consisting of 4 active GWs without deploying the SGD technique, i.e. without redundant GWs. This serves as a reference for judging both the technical need and cost effectiveness of deploying the SGD technique. The network availability is assessed for each of the 4 GWs utilising 5 dB and 10 dB of FM (as there is no switching involved), respectively.

For the selection of the 4 GWs out of 6 GWs there are 12 combinations provided that each combination has 3 GWs from different climatic regions. This selection of GWs gives the highest possible values of availabilities bearing in mind the joint exceedance time discussed earlier in Section II. Table III shows the measured network annual availability for each combination. The availability values range from 95.3497% to 96.9567% at 5dB and from 98.6269% to 99.0769% at 10dB with average values 96.1692% and 98.8586%, respectively. On average, the networks which include the GWs at Vigo and Aveiro have the worst availability despite the long distance, 173.3Km, between the two GWs compared to the distances between Chilton-Chilbolton and Athens-Lavrion.

TABLE IV
SGD PERFORMANCE OF THE 4+2 NETWORK

| Switching process Delay | 0sec | | 2sec | | 30sec | |
|---|---|---|---|---|---|---|
| SST | 5dB | 10dB | 5dB | 10dB | 5dB | 10dB |
| Network Availability (%) | 99.9645 | 99.9992 | 99.9591 | 99.9971 | 99.8841 | 99.9687 |
| Network Fade Outage (minutes) | 166.23 | 3.88 | 165.83 | 3.88 | 158.32 | 3.80 |
| Network Switching Outage (minutes) | N/A | N/A | 26.07 | 9.63 | 385 | 143 |
| Network Total Outage (minutes) | 166.23 | 3.88 | 191.9 | 13.51 | 543.32 | 146.8 |
| Number of Network Fades | 514 | 21 | 502 | 21 | 404 | 19 |
| Number of days with Outages | 20 | 2 | 20 | 2 | 20 | 2 |
| Network Mean Fade Duration (sec) | 19.4 | 11.1 | 19.8 | 11.1 | 23.5 | 12 |
| Number of Switches | 813 | 280 | 782 | 290 | 770 | 287 |
| Number of days with Switches | 175 | 110 | 175 | 115 | 176 | 110 |

In conclusion, regardless of the combination of GWs, the achieved availability cannot reach the demanding availability requirements of feeder links (e.g. 99.99%) and, consequently, the HTS systems cannot meet their service specifications.

*B. (4+2) Without Switching Process Delay*

The network of this subsection consists of 4 Active and 2 Standby stations. The choice of initial combination of Active/Standby GWs makes no difference. Further, the random selection of the Active GW from the available standby GWs ensures the unbiased statistical estimation of the number of switches. Also, the network includes 3 pairs of GWs with each pair within the same climatic area. Therefore, the results are considered a worst-case scenario (see Section II).

The measured SGD performance is listed in Table IV. For SST=5dB the feeder network annual availability is 99.9645% which is achieved via 813 ideal switches (without a switching process delay) on 175 days out of the 365 days of observation. The daily network availability is 100% except for 20 days for which the availability values range from 99.1273% to 99.9965%. For SST =10 dB the network availability is 99.9992% which is achieved via 280 ideal switches on 110 days. All days have daily network availability 100% expect for two days with availabilities 99.7442% and 99.9857%. Therefore, thanks to the SGD technique, there is a significant increase from 96.1692% to 99.9645% and from 98.8586% to 99.9992% of the feeder network availability at SST=5dB and SST=10dB, respectively.

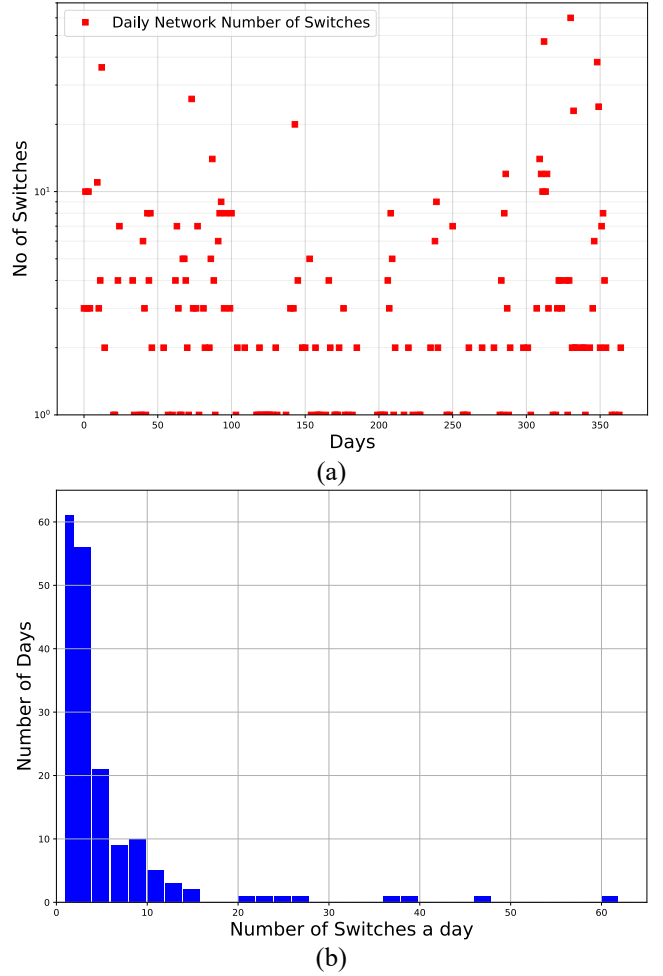

Fig. 5. Daily number of network switches for the 4+2 network with SST=5dB. a) number of switches per day b) distribution of the number of days with switches classified according to the number of switches occurred during the day.

Fig. 5a and Fig. 6a show the daily distribution of switches for SST=5dB and SST=10dB respectively. The frequency table of the number of switches per day in the network, i.e. the number of days with 1 switch or number of switches within [2,4),[4,6), …[38,40),… are shown in Fig. 5b and Fig. 6b. The number of days with more than 4 switches a day drops steeply with the increasing number of switches per day. This is more apparent as the SST increases.

The number of switches for each GW for the whole observation period is given in Table V. It seems that the number of switches of each GW is related to the climatic area where it is located. As was mentioned in Section II, the lowest numbers of fades were observed in Athens and Lavrion along with the longest on average duration compared to the other locations. This results in the smaller number of switches for the GWs at these two locations.

Fig. 7 shows the measured annual network availability and the required number of switches versus the SST, which ranges from 5dB to 14dB. There is a clear increase of the feeder network availability and decrease of the number of switches respectively as the SST increases. However, beyond 10dB there is a saturation to almost 100% availability whereas the number of switches continues to decrease. This is exactly the type of plot that allows a ground network operator to decide how to trade between the cost of the individual GW (expressed by the SST) and the number of GWs.

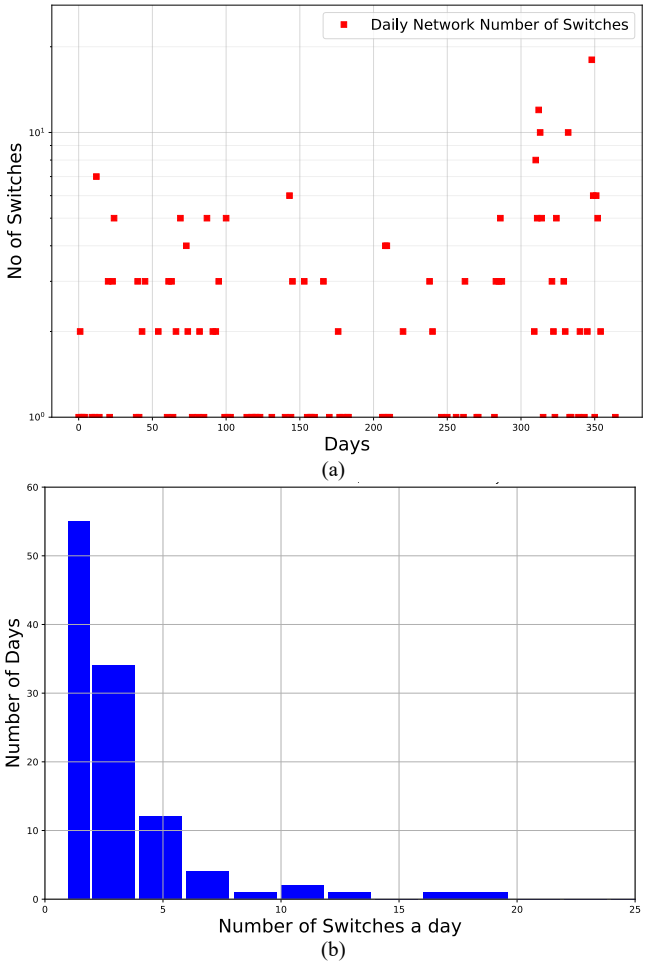

Fig. 6. Daily number of network switches for the 4+2 network with SST=10dB. a) number of switches per day b) distribution of the number of days with switches classified according to the number of switches occurred during the day.

TABLE V
NUMBER OF SWITCHES OF THE 4+2 NETWORK

| Location | Number of Switches | |
|---|---|---|
| | SST=5dB | SST=10dB |
| Chilton | 338 | 103 |
| Chilbolton | 358 | 118 |
| Athens | 203 | 50 |
| Vigo | 346 | 146 |
| Lavrion | 126 | 50 |
| Aveiro | 255 | 93 |
| Total Network switches | 813 | 280 |

For a significant fraction of the observation time, (a year in this paper), when a GW is on standby mode the attenuation on the feeder link is less that the SST. This means that the propagation impairments could be compensated by the GW's FM and the GW could be operational. This time spans from 19.90% to 39.77% and 16.03% to 44.44% at SST=5dB and SST=10dB respectively.

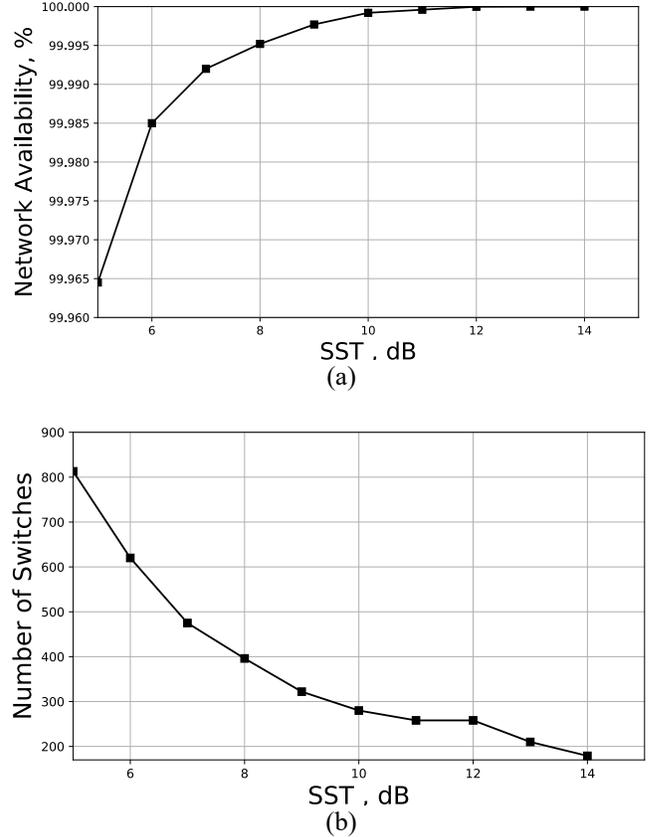

Fig. 7. Network availability (a) and number of required switches (b) versus the SST for the (4+2) network.

### C. (2+1) Without Switching Process Delay

Next, the 4+2 feeder network, i.e. 6 GWs, is split into two sub-networks each containing (2+1) GWs. The (2+1) feeder networks are structured considering the propagation dependence of the locations of the GWs and are classified into two categories: the category 1 where all the three GW pairs of the network are from different climatic regions and the category 2 where the network includes a GW pair from the same climatic area. The latter configuration may be representation of a scenario where there is a large number of GWs that can no longer be distributed in remote enough location to claim perfect weather decorrelation. However, not only the clustering of the total population of GWs into sub-networks, but also the pairing of the GWs within each sub-network may be dictated by the satellite payload connectivity between feeder beams and user beams. For example, the sites in the same climatic region may be too close to each other to be simultaneously active and still allow an adequate uplink carrier-to-co-channel interference ratio.

For the given 4+2 network there are 4 pairs of sub-networks with each containing (2+1) GWs from category 1 and 6 pairs of sub-networks containing (2+1) GWs from category 2. The performance evaluation of each pair of sub-networks is given in terms of:

i) the availability and number of switches for the individual (2+1) networks of the pair and

TABLE VI
AVAILABILITY AND NUMBER OF SWITCHES FOR (2+1) NETWORKS OF CATEGORY1 OPERATING WITHOUT SWITCHING DELAY

| Pairs of 2+1 sub-networks | GW Locations | SST=5dB | | SST=10dB | |
|---|---|---|---|---|---|
| | | Availability (%) | Number of Switches | Availability (%) | Number of Switches |
| 1 | Chilton-Vigo-Athens | 99.9439 | 420 | 99.9952 | 149 |
| | Chilbolton-Aveiro-Lavrion | 99.9660 | 351 | 99.9991 | 114 |
| | Pair performance | 99.9550 | 771 | 99.9972 | 263 |
| 2 | Chilton-Vigo-Lavrion | 99.9493 | 362 | 99.9951 | 145 |
| | Chilbolton-Aveiro-Athens | 99.9595 | 406 | 99.9971 | 127 |
| | Pair performance | 99.9544 | 768 | 99.9961 | 272 |
| 3 | Chilton-Aveiro-Athens | 99.9682 | 337 | 99.9964 | 122 |
| | Chilbolton-Vigo-Lavrion | 99.9199 | 419 | 99.9910 | 162 |
| | Pair performance | 99.9441 | 756 | 99.9937 | 284 |
| 4 | Chilton-Aveiro-Lavrion | 99.9748 | 273 | 99.9979 | 108 |
| | Chilbolton-Vigo-Athens | 99.9144 | 474 | 99.9911 | 162 |
| | Pair performance | 99.9446 | 747 | 99.9945 | 270 |

TABLE VII
AVAILABILITY AND NUMBER OF SWITCHES FOR (2+1) NETWORKS OF CATEGORY2 OPERATING WITHOUT SWITCHING DELAY

| Pairs of 2+1 sub-networks | GW Locations | SST=5dB | | SST=10dB | |
|---|---|---|---|---|---|
| | | Availability (%) | Number of Switches | Availability (%) | Number of Switches |
| 1 | Chilton-Chilbolton-Vigo | 99.6769 | 976 | 99.9606 | 326 |
| | Athens-Lavrion-Aveiro | 99.8471 | 292 | 99.9787 | 131 |
| | Pair performance | 99.762 | 1268 | 99.96965 | 457 |
| 2 | Chilton-Chilbolton-Aveiro | 99.7237 | 917 | 99.9693 | 275 |
| | Athens-Lavrion-Vigo | 99.8617 | 278 | 99.9797 | 128 |
| | Pair performance | 99.7927 | 1195 | 99.9745 | 403 |
| 3 | Chilton-Chilbolton-Athens | 99.7644 | 742 | 99.9716 | 237 |
| | Vigo-Aveiro-Lavrion | 99.8374 | 858 | 99.9853 | 401 |
| | Pair performance | 99.8009 | 1600 | 99.97845 | 638 |
| 4 | Chilton-Chilbolton-Lavrion | 99.7687 | 685 | 99.9731 | 236 |
| | Vigo-Aveiro-Athens | 99.8383 | 884 | 99.9849 | 406 |
| | Pair performance | 99.8035 | 1569 | 99.979 | 642 |
| 5 | Vigo-Aveiro-Chilton | 99.8069 | 1056 | 99.9814 | 460 |
| | Athens-Lavrion-Chilbolton | 99.8648 | 301 | 99.9814 | 117 |
| | Pair performance | 99.83585 | 1357 | 99.9814 | 577 |
| 6 | Vigo-Aveiro-Chilbolton | 99.7755 | 1145 | 99.9781 | 467 |
| | Athens-Lavrion-Chilton | 99.8678 | 309 | 99.9813 | 117 |
| | Pair performance | 99.82165 | 1454 | 99.9797 | 584 |

ii) the pair availability and the sum of the number of switches of the two (2+1) networks. The pair availability is defined as the average of the two (2+1) sub-networks availabilities.

Tables VI and VII reveal the measured SGD performance of the 4 pairs and 6 pairs of (2+1) sub-networks from category1 and category 2 respectively, for SST=5dB and SST=10dB.

For all the sub-networks pairs the availability is less than the availability of the (4+2) network. A pair of (2+1) sub-networks has, in total, equal number of active and standby GWs, i.e. 2+2 and 1+1 respectively. However, within each sub-network there is the option of switching only to one GW compared to the option of switching to two GWs for the (4+2) network. This results in a better availability for the latter.

The listed values in Tables VI and VII indicate that the 4 pairs of networks from category 1 are statistically similar and perform much better than the sub-networks from category 2. In particular, the availability of sub-networks from category 1 is greater than 99.9% and 99.99% for SST=5dB and SST=10dB respectively. On the other hand, the sub-networks from category 2 cannot reach availabilities greater than 99.87% and 99.986% for SST=5dB and SST=10dB respectively. This is expected based on the earlier rationale about joint exceedance statistics for GW pairs.

Regarding the number of switches of each GW for the sub-networks from category 1, similar conclusions are drawn as for the (4+2) network (see Section IV.B). However, for the sub-networks from category 2, the smaller number of GW switches occurs always at the GW which is located in different climatic region with respect to the other two GWs.

### D. Impact Of Switching Process Delay

Let $w$ be the time interval required for a switching from one GW to another to be completed. The following approach is adopted in the emulations of this subsection: When a switching is initiated at the time instant $t$, the whole network is frozen during the switching process delay window $w$. The network is operational, and consequently a decision for a following switch can be made, from the time instant $t+w$ onwards. The switching delay introduces an additional network outage to fade outage, the switching outage $Out_{switch}$, which reduces further the network availability and is given by the equation:

$$Out_{switch} = N_{switches} \cdot w \qquad (1)$$

where $N_{switches}$ is the number of required switches.

The SGD performance of the (4+2) network with switching process delay w equal to 2 s and 30 s, respectively for all the GWs is shown in Table IV for SST=5dB and SST=10dB. The results can be directly compared with the SGD performance of ideal switching. The differences in the values of network fade outage, number of fades and number of switches for networks operating without and with switching process delay are rather random and due to the network freeze within the time window w. The switching characteristics, e.g. number of switches of each GW, daily distribution of switches, are similar to the switching characteristics without a processing delay discussed in Section IV.B. Similar conclusions are drawn for the impact of switching process delay on the (2+1) networks.

*E. Impact Of Frequency*

To portray the SGD technique on the forward link (uplink) and in general to evaluate the impact of the GW operational frequency on the SGD technique, the 40 GHz measured time series of attenuation are scaled-up to their corresponding uplink frequency of 50 GHz. This is of high interest because, typically, a feeder link carries much more traffic in its uplink (forward link) than in the downlink (return link). Therefore, protocol and strategies for SGD should focus on the feeder uplink. The frequency scaling is performed by using the algorithm described in ITU-R P.618 [10].

At 50GHz, the (4+2) network availability over the one-year observation period for ideal switching is 99.9932%, 99.9988% and 99.9997% for SST values equal to 10 dB, 11 dB and 12 dB respectively. The corresponding required number of switches are 454, 362 and 292. A comparison of these values with the network availability and number of switches, 99.9992% and 280, for the same (4+2) network operating at 40 GHz with SST=10 dB shows that the performance of the two (4+2) networks is similar. This can be explained by the facts that:

i) The performance of SGD technique depends on the joint exceedance time of a given attenuation threshold, i.e. SST. As the frequency increases the GW links are more prone to propagation impairments. However, the strong dependence of the joint exceedance time on the spatial distribution of atmospheric phenomena significantly reduces the propagation impairments particularly if the GWs are located in different climatic regions (see section II). [11]
ii) In addition, the increase of antenna gain with the increasing frequency might allow higher levels of SST with the same antenna which reduces further the joint exceedance time. For example, the performance of the feeder network operating at 40 GHz with SST=10dB is comparable to the one operating at 50 GHz with SST=12dB.

Note, however, that in a practical SGD system, when a switching takes places this is done for both the forward and the return links.

## V. Conclusions

The performance of 4 active + 2 redundant SGD technique is assessed based on one-year measurements of the Q-band (40 GHz) propagation signal transmitted by the Aldo Paraboni payload (TDP5) of the Alphasat satellite in 6 European locations. This is a unique type of emulations given the wealth and quality of data from such a large number of locations.

It is found that SGD significantly improves the network availability but at the expense of the additional GWs and required number of switches. There is a steep increase of the feeder availability and decrease of number of switches, respectively as the SST increases. However, beyond 10 dB there is a saturation effect evident to almost 100% availability whereas the number of switches continues to decrease. A ground segment network design needs to take into account this information to balance the network cost between individual GW and network sizing.

For a significant fraction of the observation time, spanning on average from 16% to 45%, each GW is on standby mode with the attenuation on the feeder link less than SST. This means that during this time the propagation impairments could be compensated by the GW's FM and the GW could be operational. This fact maybe lends itself to considering re-using GWs among multiple satellite networks.

When the (4+2) network is split into two (2+1) networks, the sub-networks pair has always lesser availability than the (4+2) network.

As the switching from one GW to another cannot be ideal, i.e. without switching process delay, the network availability is reduced further by the number of switches times the switching process delay. The number of switches is a critical parameter for the implementation of the SGD, not only to reduce the switching outage, but also due to practical constraints of the system (e.g. maximum allowed number of switches a day, time spacing between consecutive switches). However, the required number of switches can be reduced by an optimum switching management scheme, e.g. short-term prediction of the radio channel, long-term planning of switching, selection of standby GW, selection of SST. Therefore, research activities are proposed in this particular topic to secure the cost and technically effective implementation of the SGD technique.

Finally, for the impact of operation frequency on the SGD performance it seems that system dimensioning for the application of SGD technique to higher frequencies is not proportional to severity of propagation impairments at these frequencies.

## Acknowledgment

The authors would like to warmly thank Prof. F.Perez Fontan, Prof. A.Rocha and Prof. A.D. Panagopoulos for the high quality data from Vigo, Aveiro and Greece required for the emulations. They also wish to thank Dr. A. Martellucci from ESA/ESTEC for his valuable advise.